\documentclass[12pt]{article}
 
\usepackage{epsf}
\usepackage{graphicx}
\usepackage{rotating}
\usepackage{epsfig}
\newcommand{\be}{\begin{eqnarray}}
\newcommand{\ee}{\end{eqnarray}}

\makeindex

\begin{document}

\title{Strongly Interacting Matter at High Energy Density\footnote{Lecture delivered at
the International School of Subnuclear Physics, Erice, Italy 2008}}

\author{Larry McLerran\\
  {\small\it Riken Brookhaven Center and Physics Department}\\
  {\small\it Brookhaven National
  Laboratory, Upton, NY 11973 USA}
}

\maketitle

\begin{abstract}
This lecture concerns the properties of strongly interacting matter (which is described by Quantum Chromodynamics) at very high energy density.  I review the properties of matter at high temperature, discussing the deconfinement phase transition .  At high baryon density and low temperature, large $N_c$ arguments are developed which suggest that high baryonic density matter is a third form of matter, Quarkyonic Matter, that is distinct from confined hadronic matter and deconfined matter.  I finally discuss the Color Glass Condensate which controls the high energy limit of QCD, and forms the low x part of a hadron wavefunction.  The Glasma is introduced as matter formed by the Color Glass Condensate which eventually thermalizes into a Quark Gluon Plasma.
 \end{abstract}

\section{Introduction}

This lecture is about the structure of matter at very high energy density.  Matter in thermal equilibrium
is characterized by a temperature and various chemical potentials corresponding to conserved 
charges.\cite{itoh}-\cite{pisarski1} The only conserved charge I consider is baryon number.  In addition to matter in thermal equilibrium, there is the matter which composes the dominant part of the hadron wavefunction at high energy.  This matter is a highly coherent configuration of gluons at large energy density and is called the Color Glass Condensate.\cite{cgc1}-\cite{cgc3}
 Such matter controls the interactions of strongly interacting particles in the high energy limit.  In collisions, sheets of Colored Glass pass through one another, and in the first instants after
the collisions, color electric and magnetic fields are produced which have a different topology
from the Color Glass Condensate fields.  The CGC fields are in the plane perpendicular to the direction of motion of the hadrons, but the Glasma fields are longitudinal and along the direction of motion.  Such fields bear a close resemblance to the fields hypothesized within the Lund model. \cite{lund}-\cite{lappi}

 \begin{figure}[ht]
 \begin{center}
        \includegraphics[width=0.90\textwidth]{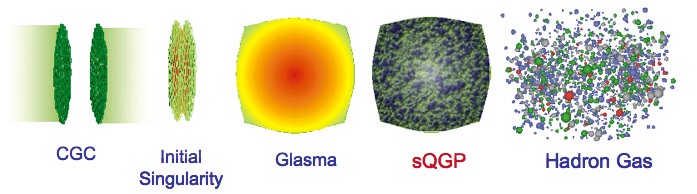}
 \end{center}       
        \caption{A schematic picture of the evolution of matter produced in the heavy ion collisions. }
\label{lilbang}
\end{figure}

A cartoon of how such high energy density matter appears in high energy nuclear collisions in shown in Fig. \ref{lilbang}.  Sheets of Color Glass Condensate collide.  When they pass through one another, the CGC fields rapidly change and a Glasma is formed.  In the limit $E \rightarrow \infty$, there is an initial singularity corresponding to this rapid change. The decay products of the Glasma matter presumably thermalize, forming a strongly interacting Quark Gluon Plasma.  The sQGP after some time hadronizes, and these hadrons make their way to detectors.  The experimental study of such collisions involves knowledge of the properties of all of the forms of matter described above.

\section{Matter at Finite Temperature}

\subsection{Confinement}

The partition function is
\begin{equation}
 Z = Tr~e^{-\beta H + \beta \mu_B N_B}
\end{equation}
where the temperature is $T = 1/\beta$ and $N_B$ is the baryon number and $\mu_B$ is the baryon number chemical potential.  Operator expectation values are
\begin{equation}
 <O> = {{Tr~ O ~e^{-\beta H + \beta \mu_B N_B}} \over Z}
\end{equation}
 Under the substitution $e^{-\beta H} \rightarrow e^{-itH}$, the partition function becomes the time
 evolution operator of QCD.  Therefore, if we change $t \rightarrow it$,and redefine zeroth
 components of fields by  appropriate factors of i, and introduce Euclidean gamma matrices with anti-commutation relations
 \begin{equation}
   \{ \gamma^\mu, \gamma^\nu \} = -2 \delta^{\mu \nu}
 \end{equation}
  then for QCD, the partition function has the path integral representation 
  \begin{equation}
  Z = \int~[dA] [d\overline \psi ] [d\psi] exp\left\{ -\int_0^\beta~ d^4x~\left( {1 \over 4 }F^2 
  +\overline \psi \left[ {1 \over i} \gamma \cdot D + m+ i \mu_Q \gamma^0 \psi \right]\right) \right\}
  \end{equation}
Here the fermion field is a quark field so that the baryon number chemical potential is
\begin{equation}
  \mu_Q = {1 \over N_c} \mu_B
\end{equation}
This path integral is in Euclidean space and is computable using Monte Carlo methods when
the quark chemical potential vanishes.  If the quark chemical potential is non-zero, various contributions appear with different sign, and the Monte Carlo integrations are poorly convergent.  Boundary conditions
on the fields must be specified on account of the finite length of the integration in time.  They are periodic for Bosons and anti-periodic for Fermions, and follow from the trace in the definition of the partition function.

\begin{figure}[ht]
        \begin{center}
        \includegraphics[width=0.50\textwidth]{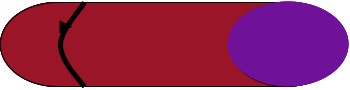}
        \end{center}
        \caption{The contour in the t plane which defines the Polyakov loop.  The space is closed in time because of the periodic boundary conditions imposed by the definition of the partition function.}
\label{line}
\end{figure}

A straightforward way to probe the confining properties of the QCD matter is to introduce a heavy
test quark.  If the free energy of the heavy test quark is infinite, then there is confinement,
and if it is finite there is deconfinement.  One can prove that the free energy of an quark added to the system is
\begin{equation}
  e^{-\beta F_q} = <L>
  \label{L}
\end{equation}
where
 \begin{equation}
    L(\vec{x}) = {1 \over N_c} Tr~ P~ e^{i \int ~dt~ A^0(\vec{x},t)}
 \end{equation}
So confinement means $<L> = 0$ and deconfinement means that $<L>$ is finite.  The path ordered phase integration which defines the line operator $L$ is shown in Fig. \ref{line}  Such a path ordered phase  is called a Polyakov loop or Wilson line.

{\bf Exercise}

{\it  Prove that the free energy of a heavy static quark added to the system is given by Eqn. \ref{L}.
To do this, use the heavy quark action as}
\begin{equation}
 S_{HQ} = \int ~dt ~\overline \psi (\vec{x},t) ~{1 \over i}\gamma^0 D^0~ 
\psi (\vec{x},t).
\end{equation}
{\it  Then show that the gluonic contribution to the action is invariant under gauge transformations that are periodic up to an element of the center of the gauge group.  The center of the gauge group is a set of diagonal matrices matrix $Z_p = e^{2\pi i p/N} \overline I$ where $\overline I$ is an identity matrix.
Show that the quark contribution to the action is not invariant, and show that $L \rightarrow Z_p L$ under this transformation.  Show that in a theory with only dynamical gluons, that the energy of a system of $n$ quarks minus antiquarks is invariant under the center symmetry transformation only if $n$ is an integer multiple of $N$.  Therefore, when the center symmetry is realized, the only states of finite free energy
are baryons plus color singlet mesons.}

In the exercise, you have shown that the realization of the center symmetry, $L \rightarrow Z_p L$
is equivalent to confinement.  This symmetry is like the global rotational symmetry of a spin system, and it may be either realized or broken.   At large separations, the correlation of a line and its adjoint, corresponding to a quark-antiquark pair is
\begin{equation}
        lim_{r \rightarrow \infty} <L(r) L^\dagger (0)> = Ce^{-\kappa r} + <L(0)><L^\dagger (0)>
\end{equation}        
since upon subtracting a mean field term, correlation functions should vanish exponentially.  Since
\begin{equation}
   e^{-\beta F_{q \overline q} (r)} = <L(r) L^\dagger (0)>
\end{equation}
we see that in the confined phase, where $<L> = 0$, the potential is linear, but in the unconfined phase,
where $<L>$ is non-zero, the potential goes to a constant at large separations.

The analogy with a spin system is useful.  For the spin system corresponding to QCD
without dynamical quarks,
the partition function  can be written as
\begin{equation}
 Z = \int~ [dA]~e^{- {1\over g^2} S[A]}
\end{equation}
The effective temperature of the spin system associated with the gluon fields is $T_{eff} \sim g^2$.
By asymptotic freedom of the strong interactions, as real temperature gets larger, the effective temperature gets smaller.  So at large real temperature (small effective temperature) we expect an ordered system, where the $Z_N$ symmetry is broken, and there is deconfinement.  For small real temperature corresponding to large effective temperature, there is disorder or confinement.

As you showed in the exercise, the presence of dynamical fermions breaks the $Z_N$ symmetry.  This is analogous to placing a spin system in an external magnetic field.  There is no longer any symmetry associated with confinement, and the phase transition can disappear.  This is what is believed to happen in QCD for physical masses of quarks.  What was a first order phase transition for the theory in the absence of quarks becomes  a continuous change in the properties of the matter for the theory with quarks.

\begin{figure}[ht]
        \begin{center}
        \includegraphics[width=0.60\textwidth]{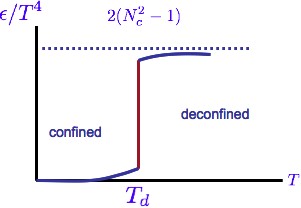}
        \end{center}
        \caption{The energy density scaled by $T^4$ for QCD in the absence of dynamical quarks.}
\label{et4}
\end{figure}

Another way to think about the confinement-decofinement transition is a change in the number of degrees of freedom.  At low temperatures, there are light meson degrees of freedom.  Since these
are confined, the number of degrees of freedom is of order one in the number of colors.  In the unconfined world, there are $2(N_c^2-1)$ gluons, and $4N_cN_f$ fermions where $N_f$ is the number of light mass fermion families.  The energy density scaled by $T^4$ is a dimensionless number and directly proportional to the number of degrees of freedom.  We expect it to have the property shown in Fig. \ref{et4} for pure QCD in the absence of quarks.  The discontinuity at the deconfinement temperature, $T_d$ is the latent heat of the phase transition.

\begin{figure}[ht]
        \begin{center}
        \includegraphics[width=0.60\textwidth]{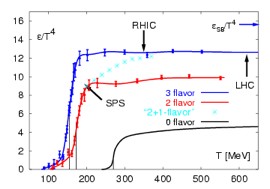}
        \end{center}
        \caption{The energy density scaled by $T^4$ measured in QCD from lattice Monte Carlo simulation.  Here there are quarks with realistic masses.}
\label{et4lat}
\end{figure}

The energy density can be computed using lattice Monte Carlo methods.  The result of such computation is shown in Fig. \ref{et4lat}.  The discontinuity present for the theory with no quarks becomes a rapid cross over when dynamical quarks are present.

The large $N_c$ limit gives some insight into the properties of high temperature matter.\cite{thooft}-\cite{thorn}  As $N_c \rightarrow \infty$, the energy density itself is an order parameter for the decofinement phase transition.  Viewed from the hadronic world, there is an amount of energy density $\sim N_c^2$ which must be inserted 
to surpass the transition temperature.  At infinite $N_c$ this cannot happen, as this involves an inifnite amount of energy.  There is a Hagedorn limiting temperature, which for finite $N_c$ would have been the deconfinement temperature.

The Hagedorn limiting temperature can be understood from the viewpoint of the hadronic world as arising from an exponentially growing density of states.  In a few paragraphs, we will argue that mesons and glueballs are very weakly interacting in the limit of large $N_c$.  Therefore, the partition function is
\begin{equation}
  Z = \int~ dm~\rho (m) e^{-m/T}
 \end{equation}
 Taking $\rho(m) \sim m^\alpha e^{\kappa m}$, so that 
 \begin{equation}
   <m> \sim {1 \over {1/T-\kappa}}
 \end{equation} 
diverges when $T \rightarrow 1/\kappa$

\subsection{A Brief Review of the Large $N_c$ Limit}

\begin{figure}[htbp]
\begin{center}
\begin{tabular} {l l l}
\includegraphics[width=0.50\textwidth] {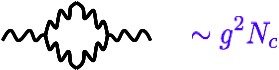}  & &
\includegraphics[width=0.46\textwidth] {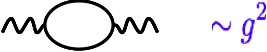} \\
& & \\
a & & b \\
\end{tabular}
\end{center}
\caption{a:  The gluon loop contribution to the heavy quark potential.  b:  The quark loop contribution to the potential}
\label{loop}
\end{figure}

The large $N_c$ limit for an interacting theory takes $N_c \rightarrow \infty $ with the 't Hooft coupling
$g^2_{'t Hooft} = g^2 N_c$ finite.  This approximation has the advantage that the interactions among quarks and gluons simplifies.  For example, at finite temperature, the disappearance of confinement
is associated with Debye screening by gluon loops, as shown in Fig. \ref{loop}a.  This diagram generates a screening mass of order $M^2_{screening} \sim g^2_{'t Hooft} T^2$.   On the other hand the quark loop contribution is smaller by a power of $N_c$ and vanishes in the large $N_c$ limit.

\begin{figure}[htbp]
\begin{center}

a  \includegraphics[width=0.75\textwidth ] {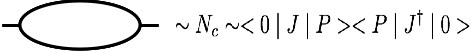}  \\
 ~ \\
~~b ~~~ \includegraphics[width=0.75\textwidth ] {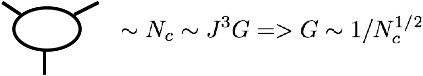} \\

\end{center}
\caption{a:  The quark loop corresponding to a current-current interaction.  b: A quark loop corresponding to a three current interaction.}
\label{int}
\end{figure}

To understand interactions consider Fig. \ref{int}a.  This corresponds to a mesonic current-current interaction through quarks.  In powers of $N_c$, it is of order $N_c$.  Gluon interactions will not change this overall factor.  The three current interaction is also of order $N_C$ as shown in Fig. \ref{int}b.  The three meson vertex, $G$ which remains after amputating the external lines, is therefore of order $1/\sqrt{N_c}$.  A similar argument shows that the four meson interaction is of order $1/N_c$.

{\bf Exercise}

{\it Show that the 3 glueball vertex is of order $1/N_c$ and the four glueball interaction of order
$1/N_c^2$}

These arguments show that QCD at large $N_c$ becomes a theory of non-interacting mesons and glueballs.  There are an infinite number of such states because excitations can never decay.  In fact, the spectrum of mesons seen in nature does look to a fair approximation like non-interacting particles.  Widths of resonances are typically of order $200 MeV$, for resonances with masses up to several $GeV$.
\subsection{Mass Generation and Chiral Symmetry Breaking}

QCD in the limit of zero quark masses has a $U(1) \times SU_L(2) \times SU_R(2)$ symmetry.  (The $U_5(1)$ symmetry is explicitly broken due to the axial anomaly.)  Since the pion field, $\overline \psi \tau^a \gamma_5 \psi$ is generated by an $SU_{L-R}(2)$ transformation of the sigma field, $\overline \psi \psi$, the energy (or potential) in the space of the pion-sigma field is degenerate under this transformation.
In nature, pions have anomalously low masses.  This is believed to be a consequence of chiral symmetry breaking, where the $\sigma $ field acquires an expectation value, and the pion fields are Goldstone bosons associated with the degeneracy of the potential under the chiral rotations.

\begin{figure}[ht]
        \begin{center}
        \includegraphics[width=0.40\textwidth]{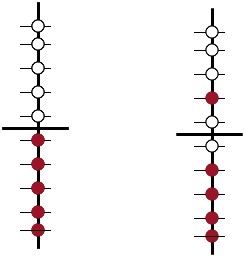}
        \end{center}
        \caption{The energy levels of the Dirac equation.  Unfilled states are open circles and filled states are solid circles.  For the free Dirac equation, negative energy states are filled and positive energy states are unoccupied, as shown on the left hand side.  A mesonic excitation corresponding to a particle hole pair is shown on the right hand side.}
\label{hole}
\end{figure}

Such symmetry breaking can occur if the energy of a particle-antiparticle pair is less than zero, as shown in Fig. \ref{hole}.  On the left of this figure is the naive vacuum where the negative energy states associated with quark are filed.  The right hand side of the figure corresponds to a particle hole excitation, corresponding to a sigma meson.  Remember that a hole in the negative energy sea corresponds to an antiparticle with the opposite momentum and energy. If the $\sigma$ meson excitation has negative energy, the system is unstable with respect to forming a condensate of these mesons.

At sufficiently high temperature, the chiral condensate might melt.  Indeed this occurs.\cite{karsch}  For QCD,
the chiral and deconfinement phase transition occur at the same temperature.  At a temperature of about $170 - 200 MeV$, both the linear potential disappears and chiral symmetry is restored.  It is difficult to make a precise statement about the indentification of the chiral and deconfinement phase transitions,
since as argued above, for QCD with quarks, there is not a real phase transition associated with deconfinement.\cite{karsch1}-\cite{fodor}  Also, when quarks have finite masses, as they do in nature, chiral symmetry is not an exact symmetry, and there need be no strict phase transition associated with its restoration.  nevertheless, the cross over is quite rapid, and the point, and there are rapid changes in the both the potential and the sigma condensate $<\overline \psi \psi >$ at temperatures which are in a narrow range.

\section{Finite Density}

\begin{figure}[ht]
        \begin{center}
        \includegraphics[width=0.50\textwidth]{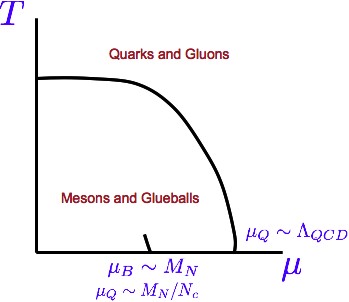}
        \end{center}
        \caption{The popular folklore for the phase diagram of QCD.}
\label{phasediagram}
\end{figure}

When the deconfinement phase transition was first discussed, the phase diagram of nuclear matter was postulated.  This postulate is shown in Fig. \ref{phasediagram}.  In such a diagram, the world is simply divided into that of a world of mesons and glueballs, corresponding to confinement, and to a deconfined world of quarks and gluons.  There is a weak liquid gas phase transitions corresponding to nuclear matter, at $\mu_B \sim M_N$ and very low temperature.  The deconfinement baryon chemical potential is assumed to be significantly larger than $M_N$.  (There may be weak phase transitions corresponding to color superconductivity, which for purposes of simplicity, I shall not discuss in this lecture.\cite{alford})

The problem with the conventional wisdom about the phase diagram of QCD is that it is not consistent with our large $N_c$ understanding.\cite{pisarski1}  As shown in Fig. \ref{loop}, the quark loop contribution to the
quark-antiquark potential is suppressed by a power of $N_C$ relative to that of the gluons, and vanishes in the large $N_c$ limit at fixed value of quark chemical potential, $\mu_Q \sim \Lambda_{QCD}$.  It is only through such loops that the effects of finite baryon number density enter into the thermodynamic potential, or correct the quark-antiquark potential.  The effects of quark loops on the quark-antiquark potential can only become important when $\mu_Q \sim \sqrt{N_c}\Lambda_{QCD}$.  This occurs
when the density of baryons is $\rho_B \sim N_C^{3/2} \Lambda^3_{QCD}$ which is parametrically large compared to the scale at which the coupling should become weak, $\rho_B \sim \Lambda^3_{QCD}$,
and where we would have naively expected deconfinement to occur.

\begin{figure}[ht]
        \begin{center}
        \includegraphics[width=0.50\textwidth]{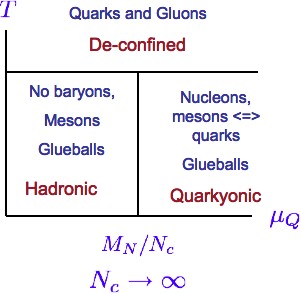}
        \end{center}
        \caption{The phase diagram of QCD at large $N_c$.}
\label{largen}
\end{figure}

At large $N_C$, there is another order parameter besides the free energy of a heavy test quark.  The baryon number density itself is an order parameter.  At finite chemical potential,
\begin{equation}
 \rho_B \sim e^{\mu_B/T - M_B/T} \sim e^{-N_c\{M_B/N_cT - \mu_Q/T \}}
\end{equation}
For $m_Q \le M_B/N_c$, the baryon number is always zero in the large $N_c$ limit.  Above it, there is a finite baryon number density and confinement.  Correspondingly, if one increases the temperature,
at some temperature independent of $\mu_Q$, there is deconfinement.  Such a phase diagram is shown in Fig. \ref{largen}

We call the third phase of matter, which has finite baryon number density but is confined, Quarkyonic.  The name expresses the fact that the matter is composed of confined baryons yet behaves like quarks
at high densities.   At chemical potentials parametrically large compare to $\Lambda_{QCD}$ but small compared to $\sqrt{N_c} \Lambda_{QCD}$, the bulk properties of the matter are computable in perturbation theory.  Therefore the energy density should scale as $N_c$.  The quarks should be approximately chirally symmetric.  There may be non-perturbative effects associated with confinement and chiral symmetry restoration near the fermi surface, since there the interactions are sensitive to long distance effects, but the bulk properties should look like almost free quarks.

There is another order parameter for the system, the energy density itself.  In the hadronic phase, $\epsilon \sim O(1)$.  In the de-confined phase, gluons dominate the number of degrees of freedom, $\epsilon \sim N_c^2$.  The Quarkyonic phase is quarks plus thermal excitations of glueballs and mesons, so that $\epsilon \sim N_c$

The transition between hadronic matter and Quarkyoninc matter is in a narrow range of quark chemical potential.  The typical Fermi momentum for the transition is $k_F \sim \Lambda_{QCD}$.  On the other hand, the quark fermi energy is
\begin{equation}
  \mu_Q \sim(M_N + k_F^2/2M_N)/N_c \sim \Lambda_{QCD}(1 + O(1/N_c^2))
\end{equation}

Since the quarks are confined, there should be a dual description of the matter at large $N_c$ in terms of either quarks or baryons.  A baryonic description is provided by the Skyrme model.\cite{adkins}  In large $N_c$, the Skyrme action is of order $N_c$, which agrees parametrically with the dependence on $N_c$ of a gas of weakly coupled quarks, and agrees with the results above.

In the discussion above, I have implicitly assumed that the limit $N_c \rightarrow \infty$ is taken while the number of quark flavors $N_f$ is held fixed.  If the number of flavors is large, then there there is never
a linear potential between quarks because quark pairs can be made from the vacuum that short out the potential.  In the fixed $N_f$ limit, it is known from lattice Monte-Carlo studies that the linear potential provides a good description of the quark-antiquark interaction for distances between $.3~Fm \le r \le 2~Fm$.    This suggests that the fixed $N_f$ large $N_c$ limit is valid for some range of distance scales.

For $N_F \sim N_c \rightarrow \infty$, there is a high degeneracy of low mass baryonic states.  The number of lowest mass baryons with $I = J = 1/2$ can be computed with a result $e^{N_cF(N_f/N_c)}$.
The baryon number density remains a good order parameter, and $\rho_B \sim e^{N_cF(N_f/N_c) + \mu_B/T - M_B/T}$.  For $T F(N_f/N_c) + \mu_Q < M_B/N_c$ there is a baryon number zero phase and for greater values thee is a finite baryon number.  The transition is caused by the Hagedorn spectrum
of baryonic resonances.

Presumably for QCD, there are reasonably sharp cross overs or phase transitions that separate the hadronic, quarkyonic and confined worlds.  The energy density of the confined world is $O(N_f)$,
the Quarkyonic is $\sim 2N_cN_f$, and the deconfined is $\sim 4N_cN_f +2N_c^2$.  Precisely how chiral symmetry restoration appears in this phase diagram is not yet known.

There is a hint from heavy ion collisions about Quarkyonic matter.  If one measures the abundances of various species of particles produced in heavy ion collisions, one may fit these numbers to a temperature and baryon number chemical potential.\cite{redlich}  These temperatures and chemical potentials are interpreted as those for which the system freezes out or decouples.  At some low temperature and density the matter is expanding so rapidly that the matter cannot maintain thermal equilibrium by rescattering.  It is expected that such a freeze out temperature reflects the position of phase transition surfaces, since at a phase transition, the number of degrees of freedom change very rapidly.  This results in a rapid lowering of the density as one goes across the transition surface.  The lower density makes it harder for the system to maintain itself in thermal equilibrium.  In numercal simulations of the properties of heavy ion collisions, this seems to be a reasonable first approximation.
\begin{figure}[ht]
        \begin{center}
        \includegraphics[width=0.50\textwidth]{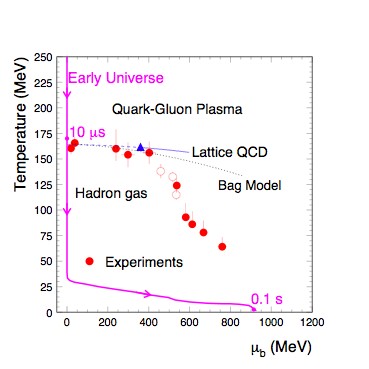}
        \end{center}
        \caption{Freeze out temperature and baryon chemical potential determined by heavy ion collisions.}
\label{redlich}
\end{figure}

In Fig. \ref{redlich}, the freeze out temperature and baryon chemical potential are plotted.  At low baryon density, the curve follows expectations that the freeze out surface follows the phase boundary between the hadronic and confined world.  At a chemical potential of about $\mu_B \sim ~500 ~MeV$, the decoupling line deviates from expectations for the confinement-deconfinement surface.  This point is about where baryons begin to dominate the energy density.  The line intersects the $T = 0$ axis at about $\mu_B = M_N$, which is what we expect for the Quarkyonic phase.  If this interpretation is correct, then
experiment has given us the rough parameters of the Quarkyonic-Hadronic boundary. 

\section{Color Glass Condensate}

\begin{figure}[ht]
        \begin{center}
        \includegraphics[width=0.50\textwidth]{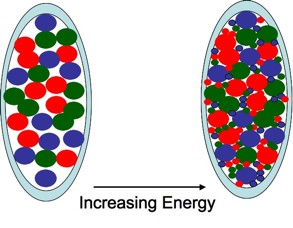}
        \end{center}
        \caption{More gluons added to a hadron as the hadron energy increases..}
\label{saturate}
\end{figure}

A remarkable experimental fact about the hadrons at high energy is that their size is very slowly increasing, but the number of gluons increases rapidly.  This means that the high energy limit is the high gluon density limit.  One would think that ultimately the density would saturate at high energy, since at some point the density would be so large that the repulsive gluon interactions would prevent further growth of the gluon number.  This is almost true.\cite{glr}-\cite{mq}  For gluons of a fixed size, the number stops growing rapidly, but one can then add in more gluons of smaller size, until the density of such gluon is high enough so that their density saturates.  This can go on forever proceeding to saturate gluons of smaller and smaller sized as the typical energy scale goes to infinity.   This is illustrated in Fig. \ref{saturate}.

This high energy density gluonic matter is the Color Glass Condensate.\cite{cgc1}-\cite{cgc3} The name color comes from the color of the gluons which compose it.  The word condensate arises from the high phase space density of the gluons.  Defining
\begin{equation}
        \rho_{PS} = {{dN} \over {dyd^2p_Td^2x_T}}
\end{equation}
where $y$ is the gluon rapidity $y = {1 \over 2} ln((E+P_z)/(E-P_z))$.  

The energy of such a distribution of gluons is
\begin{equation}
   E = -\kappa \rho_{PS} + \kappa^\prime \alpha_S \rho^2_{PS}
\end{equation}
The first term in this expression is negative because the gluon density spontaneously generates itself, and the last term is positive because of repulsive gluon interactions (needed because the energy is bounded from below).  This is extremized at $\rho_{PS} \sim1/ \alpha_S$, which is typical of condensation
phenomena in Bose condensates and superconductivity.  Such condensates are highly coherent, and the fields associated with them are classical, because of the high occupation number.  Here $\alpha_S << 1$ because the transverse density of gluons is high compared to $\Lambda^2_{QCD}$.

The word glass is in the name because the partons which make the coherent CGC field are fast moving.
Let us imagine we are in the infinite momentum frame of the hadron.  The fast moving gluons produce the more slowly moving ones.  Those at the lowest  allowed value of momentum, which  is of order $\Lambda_{QCD}$ are slow moving, and their natural time scale of evolution would be $t \sim 1/\Lambda_{QCD}$.  Because they arise from a coherent field produced from the fast partons, their time scale of evolution is Lorentz time dilated, and they evolve slowly compared to natural times scales.  This is characteristic of glassy material.

The Color Glass Condensate is made of the Lorentz boosted Couomb fields of the fast moving gluons in the hadron.  In light cone coordinates, the gluons are on a sheet of small extent in $x^- = t-z$.  They evolve slowly in $x^+ = t + z$.

{\bf Exercise}

{\it  Show that the because fields are rapidly varying in $x^-$ and slowly varying in $x^+$, that $F^{i+}$ is big, $F^{i-}$ is small, and $F^{ij}$ is of intermediated strength.  Show that the large fields satisfy
$\vec{E}  \perp \vec{B} \perp \hat{z}$, so that the color electric and magnetic fields have the same
Lorentz structure as the Lienard-Wiechart fields of electrodynamics.}
\begin{figure}[ht]
        \begin{center}
        \includegraphics[width=0.25\textwidth]{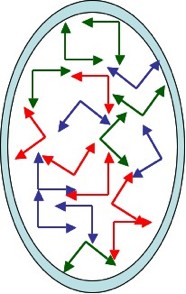}
        \end{center}
        \caption{The color electric and color magnetic fields in the Color Glass Condensate.}
\label{lwfields}
\end{figure}

The color electric and color magnetic fields of the CGC are shown in Fig. \ref{lwfields}.  These fields are of the form of the Lienard-Wiechart fields of electrodynamics.  The color orientation is random.  Their density is determined by the saturation momentum.  Integrating the phase space density up to the highest momentum scale where the gluons are still highly coherent gives\cite{nardi}
\begin{equation}
  {1 \over {\pi R^2}} {{dN} \over {dy}} = {1 \over \alpha_S} Q_{sat}^2
\end{equation}

To determine the dependence of the saturation momentum scale on energy, one needs a theory of the CGC.  This is given by QCD plus renormalization group in the gluon energy.  I have argued that gluon of higher momentum acts as a source for those of low momentum.  The scale which separates high energy gluons from soft gluons is arbitrary.  This arbitrariness is resolved by renormalization group equations.
Suppose we have the theory separated between sources and fields and there is one separation momentum scale $P_1$.  We can move to a new scale $P_2$ by integrating out fields to produce sources at the new scale.  This procedure can be done when the coupling is weak, even though the CGC fields are strong.   It turns out that the solution to this renormalization group equation is universal,
so that the high energy limit and the CGC are independent of hadron.  The solution also has the property that the saturation momentum grows forever as the energy increases.

The saturation momentum scale acts as an infrared cutoff.  For distance scales less than $1/Q_{sat}$,
fields resolve individual sources of charge.  For distance scales greater than $1/Q_{sat}$, the fields see multiple sources of charge which average to zero.  Gluonic field strengths should be softer by one power of $r$ at large distances or one power of $p$ at small momentum than one would have ignoring coherence.  For example, the gluon distribution itself, which is proportional to 
the gluon strength squared $<a^\dagger a> \sim A^2$, goes as
\begin{eqnarray}
   {{dN} \over {dyd^2p_T}} &  \sim & {1 \over {p_T^2}}  ~~~~~~~~ p_T \ge Q_{sat} \nonumber \\
   {{dN} \over {dyd^2p_T}} &  \sim & lm(p_T)   ~~~ p_T \le Q_{sat}
 \end{eqnarray}
When applied to the computation of hadron-hardon scattering, the total multiplicity becomes computable because the $1/p_T^4$ singularity of perturbation theory becomes $ \sim ln(p_T)$, which is integrable.  

\section{The Glasma}

The collision of two hadrons may be visualized as that of two very thin sheets of Colored Glass.\cite{weigert}-\cite{lappi}  The thickness of these sheets is determined by $\alpha_S \Delta y \sim 1$ since if one attempts a classical field description, quantum corrections to the classical field become important when this is satisfied.  Using that 
\begin{equation}
   y = { 1\over 2} ln ((E+p_z)/(E-p_Z)) = ln((E+p_z)/M_T) \sim -ln(\Delta z Q_{sat})
\end{equation}
This gives the thickness of the sheet as $\Delta z \sim e^{-\kappa/\alpha_S}/Q_{sat}$.   For small $\alpha_S$ this is much much smaller than the scale set by the saturation momentum.

\begin{figure}[ht]
        \begin{center}
        \includegraphics[width=0.50\textwidth]{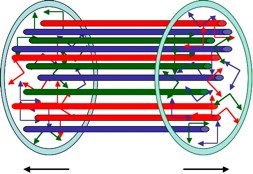}
        \end{center}
        \caption{The color electric and color magnetic fields of the Glasma immediately after the collision of two hadrons.}
\label{glasma}
\end{figure}

In the time it takes the sheets to pass through one another, the topology of the fields change.  As I shall argue in the following paragraphs, the fields which originally are embedded in the sheets of Colored Glass tsansform into longitudinal flux tubes of color electric and color magnetic fields, as shown in Fig. \ref{glasma}.
This configuration of fields is called the Glasma because it is created by the Color Glass Condensate and eventually decays into a Quark Gluon Plasma.

To understand the origin of the Glasma fields, we need to first understand a little more about the CGC fields.  The CGC fields are generated by a source of field along the lightcone, $x^\pm = 0$. For definiteness, consider the sheet with $x^- = 0$. The Yang-Mills equations may be solved by a vector potential which is a two dimensional gauge transform of vacuum fields for either of the regions 
$x^- < 0$ or $x^- > 0$.  We choose different gauge transforms of vacuum in either of these regions,
and the discontinuity in Yang-Mills equations becomes the source of color charge on the sheets.

\begin{figure}[ht]
        \begin{center}
        \includegraphics[width=0.60\textwidth]{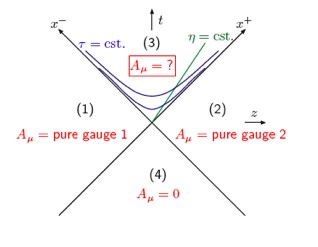}
        \end{center}
        \caption{A space-time diagram for the scattering of two sheets of colored glass}
\label{scattering}
\end{figure}

Now consider the scattering problem illustrated in Fig. \ref{scattering}.
In addition to $x^\pm$, there are the space-time rapidity variable $\eta = { 1 \over 2} ln(x^+/x^-)$,
and a proper time $\tau = \sqrt{x^+x^-}$.  

Inside the backward lightcone, I use gauge freedom to choose the field $A^\mu = 0$.  In the right hand and left hand side of the light cone I choose fields which are pure gauge transforms of vacuum.  This guarantees that along the backwards lightcone, the Yang-Mills equations are solved.  If I choose,
$A^\mu = A^\mu_1 + A^\mu_2$ infinitesimally close to the forward light cone, but inside it, then one gets the correct discontinuity to give the sources which sit on the forward light cone.  Of course, this field configuration is no longer a gauge transformation of vacuum, and will evolve.  It can be shown that this solution is independent of the longitudinal coordinate, the space-time rapidity.

The fields in the forward-lightcone are initially only longitudinal electric and magnetic fields.  This is because in the forward light cone, there are induced sources for the color electric and magnetic fields:
\begin{eqnarray}
   \vec{\nabla} \cdot \vec{E}_{1,2} &  = & \vec{A}_{2,1} \cdot \vec{E}_{1,2} \nonumber \\
    \vec{\nabla} \cdot \vec{B}_{1,2} &  = & \vec{A}_{2,1} \cdot \vec{B}_{1,2} 
\end{eqnarray}
These fields are independent of longitudinal coordinate.  They generate equal and opposite densities of 
color electric and color magnetic charge density on the sheets.  The average energy density stored in the magnetic fields is the same as that in the electric fields because of the duality of the Yang-Mills equations under $E \leftrightarrow B$, and the symmetry of the initial conditions.

The classical fields have large strength, and decay by classical evolution.  Presumably they
evolve into an interacting Quark Gluon Plasma.  There are in fact unstable modes of these solutions and
given enough time may develop turbulence and might classically thermalize the fields.\cite{mrocz}-\cite{fukushima}

Note also that these classical fields carry a non-zero Chern-Simons charge, since $E \cdot B$ is non zero  The integral of this charge vanishes, but on the transverse size scale of $1/Q_{sat}$ the charge density is non zero.  It is expected that non-zero total Chern-Simons number density will be generated by the time evolution of fluctuations, as is the case in baryogenesis in electroweak theory.\cite{kpt}-\cite{harmen}

\section{Summary}

I have presented a description of some of the properties of high energy density matter which might be made in high energy collisions.  The properties of matter in thermal equilibrium can be tested in heavy ion collisions.  The Color Glass Condensate has various phenomenological tests in the theory of high energy lepton-nucleus scattering and hadron collisions.  It has a successful phenomenology,
which will be tested with greater precision in experiments at LHC.  The concept of the Glasma is fairly new, and has qualitative features in accord with results from RHIC, and will also be tested in high energy heavy ion collisions at LHC.


\section{Acknowledgments}
I gratefully acknowledge my colleagues Yoshimasa Hidaka, Rob Piasarksi, and Raju Venugopalan
for their many discussion with me concerning the ideas presented here. 

This manuscript has been authorized under Contract No. DE-AC02-98CH0886 with the US Department of Energy.

\end{document}